\documentclass[apjl]{emulateapj}

\usepackage{epsfig}
\usepackage{amsmath, amssymb,bm}
\usepackage[varg]{txfonts}
\usepackage{color}
\usepackage{aas_macros}
\usepackage{ulem}

\usepackage[backref, hidelinks]{hyperref}
\hypersetup{breaklinks=true,colorlinks=true,urlcolor=blue, linkcolor=black,  citecolor=blue,pagecolor=red, bookmarksopen=true}

\shorttitle{Primordial giant planet obliquity}
\shortauthors{R. G. Martin \& P. J. Armitage}

\begin{document}

\title{Primordial giant planet obliquity driven by a circumplanetary disk}

\author{Rebecca G. Martin\altaffilmark{1}}
\author{Philip J. Armitage\altaffilmark{2,3}}
\affil{\altaffilmark{1}Department of Physics and Astronomy, University of Nevada, Las Vegas,
4505 South Maryland Parkway, Las Vegas, NV 89154, USA}
\affil{\altaffilmark{2}Center for Computational Astrophysics, Flatiron Institute, New York, NY 10010, USA}
\affil{\altaffilmark{3}Department of Physics and Astronomy, Stony Brook University, Stony Brook, NY 11794, USA}

\label{firstpage}
\begin{abstract} 
Detached circumplanetary disks are unstable to tilting as a result of the stellar tidal potential. We examine how a tilted circumplanetary disk  affects the evolution of the spin axis of an oblate planet. The disk is evolved using time-dependent equations for linear wave-like warp evolution, including terms representing the effect of the tidal potential and planetary oblateness. For a disk with a sufficiently large mass, we find that the planet spin quickly aligns to the misaligned disk. The tilt of the planetary spin axis then increases on the same timescale as the disk. This can be an efficient mechanism for generating primordial obliquity in giant planets. We suggest that directly imaged exoplanets at large orbital radii, where the disk mass criterion is more likely to be satisfied, could have significant obliquities due to the tilt instability of their circumplanetary disks.
\end{abstract} 
  
\keywords{ 
accretion, accretion discs -- hydrodynamics -- instabilities --planets
and satellites: formation -- planetary systems -- stars: pre-main sequence
} 
 
\section{Introduction}   
The rotation state of planets is one of the basic observables that offers information about planet formation and planetary system evolution. For giant planets, that open a gap in the protoplanetary disk \citep{LP1986}, the primordial spin state is determined by the interaction between the growing planet and its {\it circumplanetary} disk, which forms because the planet is much smaller than the Hill radius \citep{Artymowicz1996,Lubow1999,DAngelo2002}. The magnitude of the planetary angular momentum that results from circumplanetary disk accretion depends upon whether the planet-disk interaction is hydrodynamic \citep{dong21} or magnetospheric \citep{konigl91,batygin18}, but once set is expected to be conserved in the absence of tidal effects. In the Solar System Jupiter and Saturn rotate rapidly, but at a rate that is still only 30-40\% of their nominal breakup angular velocity. For a small sample of very massive extrasolar planets, \citet{bryan18} inferred somewhat lower but non-zero rotation rates.

Misalignment between the spin and orbital angular momentum vectors of planets---the planetary {\it obliquity}---is common in the Solar System. Jupiter's obliquity is low (3$^\circ$), but Saturn (27$^\circ$) and the ice giants have highly significant obliquities. Measurement of extrasolar planetary obliquities is challenging, but there are some indications of non-zero obliquity for the directly imaged planetary mass companion in the 2M0122 system \citep{bryan20}. Unlike the magnitude of the angular momentum, the obliquity is relatively fragile to late-time evolutionary changes. Resonances between the precession frequencies of planetary spin axes and orbits affect both terrestrial and giant planets in the Solar System \citep{laskar93,ward04,ward06}. A related process, where the orbital precession is driven by the gravitational potential of the protoplanetary disks, can generate early-time obliquity \citep{Millholland2019}. Giant impacts \citep{Safronov1966,Benz1989,Morbidelli2012}, and planet-planet scattering events \citep{li21,hong21}, can also lead to planet spin-orbit misalignment.

In this Letter, we investigate another mechanism that can generate primordial obliquity, as a byproduct of planetary interaction with a tilted circumplanetary disk. The physical properties of circumplanetary disks, specifically their large size relative to the Hill sphere, and moderately large geometric thickness, favor growth of the tilt instability of disks in binary systems \citep{Lubow1992,Lubow2000,MZA2020}. The tilt may be able to grow large enough to excite Kodai--Lidov oscillations \citep{Kozai1962,Lidov1962}, at which point eccentricity is induced and accretion becomes episodic \citep{Martinetal2014,Martin2021}. Here, we show that a circumplanetary disk, whose tilt is being excited by an external tidal potential, can couple to the planetary spin via oblateness, and generate obliquity. The planet exerts a torque on the disk \citep[e.g][]{Tremaine2014,Speedie2020} and an equal and opposite torque is felt by the planet. This can change the spin axis of the planet depending upon the ratio of the disk and planet spin angular momenta. An analogous process can realign a spinning black hole with a misaligned disk  \citep{SF1996,Martinetal2007,Perego2009}.  We show that the torque from the planet causes the planet spin and the disk to precess around the sum of their angular momenta.  They align towards each other on a relatively short timescale. On a longer timescale, both can have increasing tilts as a result of the tidal tilt instability, if the disk mass is sufficiently large.

\section{Circumplanetary disk and planet-spin model}

We model a circumplanetary disk around a planet of mass $M_{\rm p}$ that orbits a star of mass $M_{\rm s}$ at orbital separation $a_{\rm p}$. The angular velocity of the planet-star system is $\bm{\Omega}_{\rm p}=\Omega_{\rm p} \bm{e}_z$ where the angular frequency is $\Omega_{\rm p}=\sqrt{G (M_{\rm s}+M_{\rm p})/a_{\rm p}^3}$. The disk has surface density $\Sigma(r)$ (that is fixed in time) and a unit vector to describe the tilt at each radius, $\bm{l}_{\rm d}(r,t)=(l_{{\rm d},x},l_{{\rm d},y},l_{{\rm d},z})$. The tilt is assumed to be small so that $l_{{\rm d},z}\approx 1$ and $l_{{\rm d},x}$, $l_{{\rm d},y} \ll 1$. In order to explore the evolution to high inclination we would need to use hydrodynamical simulations.

A circumplanetary disk is expected to be in the wave-like disk regime \citep{PP1983} since the \cite{SS1973} $\alpha$ viscosity parameter is much smaller than the disk aspect ratio $H/r$ \citep[e.g.][]{Martin2019warp}, where $H$ is the disc scale height and $r$ is the spherical distance from the planet. Thus, we solve the 1D wave-like warped disk equations in the frame of the planet-star system. The disk is in Keplerian rotation around the planet at all radii with angular frequency given by $\Omega=\sqrt{G M_{\rm p}/r^3}$.  We solve equations~(12) and~(13) in \cite{Lubow2000} with an additional torque on the disk from the planet. These can be written as
\begin{equation} 
\Sigma r^2 \Omega  \left(  \frac{\partial \bm{l}_{\rm d}}{\partial t}+\bm{\Omega_{\rm p}}\times \bm{l}_{\rm d}\right)
=\frac{1}{r}\frac{\partial \bm{G}}{\partial r} +{\bm T}_{\rm s}+{\bm{T}_{\rm p}}
\label{eql}
\end{equation}
and
\begin{equation}
    \frac{\partial \bm G}{\partial t}+\bm{\Omega_{\rm p}}\times \bm{G}+\alpha \Omega \bm{G} =\frac{\Sigma H^2r^3 \Omega^3}{4}\frac{\partial \bm{l}_{\rm d}}{\partial r},
    \label{eqG}
\end{equation}
where $2\pi \bm{G}$ is the internal disk torque. 
We only need to solve the first two components of these vector equations since the disc angular momentum vector is a unit vector. We use the boundary conditions $\bm{G}=0$ and $\partial \bm{l}_{\rm d}/\partial r=0$ at $r=r_{\rm in}$ and $r=r_{\rm out}$. The initial conditions are chosen to be $\bm{G}=0$ and the disk is initially flat but tilted by $10^\circ$ with $l_{{\rm d},x}=\sin(10^\circ)$ and $l_{{\rm d},y}=0$.

We work in a frame where  the star is on the positive $x$ axis and so the stellar torque on the disk per unit area  is $\bm{T}_{\rm s}=(T_{{\rm s},x},T_{{\rm s},y})$ where $T_{{\rm s},x}=0$ and
\begin{equation}
    {T}_{\rm s,y} =-\frac{3GM_{\rm s}}{2a_{\rm p}^3}\Sigma r^2 l_{{\rm d},x}.
\end{equation}
The torque per unit area from a spinning oblate planet on the disk is given in linear theory by
\begin{equation}
    T_{{\rm p},x}=T(l_{{\rm d},y}-l_{{\rm p},y})  
    \end{equation}
    and
    \begin{equation}
   T_{{\rm p},y}= T(-l_{{\rm d},x}+l_{{\rm p},x}) 
\end{equation}
where
\begin{equation}
    T=\frac{3 G M_{\rm p}r_{\rm p}^2J_2}{2 r^3}   \Sigma
\end{equation}
\citep[e.g.][]{Tremaine2014,Speedie2020}. Here $r_{\rm p}$ is the radius of the planet and $J_2$ is the quadrupole gravitational harmonic.  In the solar system, the giant planets Jupiter, Saturn, Uranus and Neptune have $J_2=0.015, 0.016, 0.003$ and $0.004$ respectively.

The planet spin angular momentum vector is $J_{\rm planet} \bm{l}_{\rm p}$, where $\bm{l}_{\rm p}$ is a unit vector and 
\begin{equation}
    J_{\rm planet} = \frac{2\pi k M_{\rm p}r_{\rm p}^2}{P_{\rm rot}},
    \label{eqa}
\end{equation}
where $k=0.205$ (appropriate if the planet is a polytrope with $n=1.5$).
Over the timescales considered in this work,  we ignore the effects of accretion on to the planet and assume that the magnitude of the planet spin angular momentum does not change. The direction of the spin changes as a result of the torque from the disk on the planet according to
\begin{equation}
    J_{\rm planet} \left(\frac{d\bm{l_{\rm p}}}{dt}+\bm{\Omega}_{\rm p}\times \bm{l}_{\rm p}\right)=-\int_{r_{\rm in}}^{r_{\rm out}} 2 \pi r \bm{T}_{\rm p} \, dr   .
    \label{eqA}
\end{equation}
We take the initial condition that the planet spin is aligned with its orbit with $\bm{l}_{\rm p}=\bm{0}$. 

We solve the coupled integro-differential equations~(\ref{eql}),~(\ref{eqG}) and~(\ref{eqA})  as an initial value problem for $\bm{l}_{\rm d}$, $\bm{G}$ and $\bm{l}_{\rm p}$ using finite differences. The method is first order explicit in time.   The grid extends from inner radius $r_{\rm in}$ up to $r_{\rm out}$. In order to obtain convergence, we use a non-uniform grid.  There are 25 grid points  spaced linearly in $\log(r)$ from the inner radius up to $r_{\rm change}=0.02\,r_{\rm H}$. There are 75 points that are linearly distributed with radius from $r_{\rm change}$ up to $r_{\rm out}$.  

The inclinations of the disk and planet are determined via 
\begin{equation}
    i=\cos^{-1}\left(\sqrt{1-l_x^2-l_y^2}\right)
\end{equation}
and the phase angles as
\begin{equation}
    \phi=\tan^{-1}\left(\frac{l_y}{l_x}\right).
\end{equation}
For the disk, we  calculate the angular momentum weighted vector components. We transform these back into the inertial frame in our plots by rotating about the $z$ axis by $\Omega_{\rm p}t$.

\section{Disk evolution with no planetary torque}

We model a Jupiter mass and radius planet orbiting at $a_{\rm p}=5.2\,\rm au$.  The circumplanetary disk has surface density profile $\Sigma \propto r^{-3/2}$ distributed between the inner disk radius  $r_{\rm in}=2\,r_{\rm p}$ and the outer radius $r_{\rm out}=0.4\,r_{\rm H}$, where $r_{\rm H}$ is the Hill radius. The inner disk radius is close to the peak in the surface density of a steady state disk that joins on to the planet at radius $r_{\rm p}$. The outer radius is chosen to be close to the tidal truncation radius of a circumplanetary disk \citep{MartinandLubow2011}\footnote{Note that a misaligned circumplanetary disk may be tidally truncated at a larger radius than a coplanar disk \citep{Lubow2015,Miranda2015}. We do not include this correction, as our analysis formally applies only to moderate tilt values.}. We take the disk parameters $\alpha=0.01$ and $H/r=0.14$. With this disk aspect ratio the disk is strongly unstable to tilting. As shown in \cite{MZA2020}, the combination of the disk aspect ratio and the disk size determines the instability. While it appears that instability requires specific parameters in the analytic model, hydrodynamic simulations show that that a disk with a more realistic surface density profile is also unstable to tilting.  

\begin{figure} 
\begin{centering} 
\includegraphics[width=0.45\textwidth]{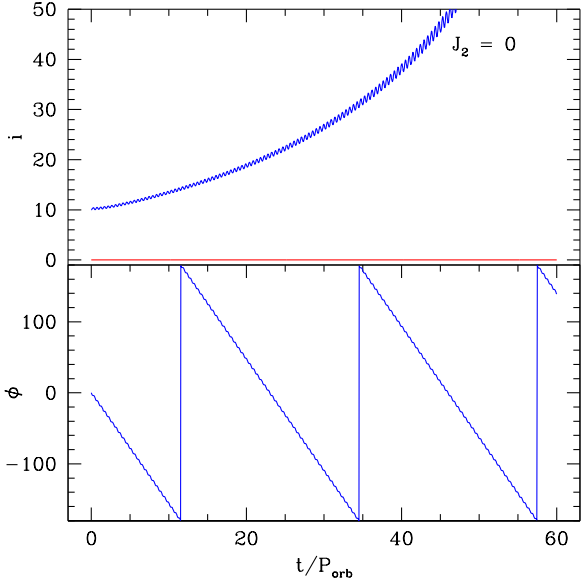}
\end{centering} 
\caption{The disk (blue) and planet (red)  inclination (upper panel) and nodal phase angle (lower panel) evolution with no planet torque, $J_2=0$. }
\label{j0} 
\end{figure}

Figure~\ref{j0} shows the evolution of the inclination and nodal phase angle for a circumplanetary disk in the limit where there is no planet torque, $J_2=0$. The blue lines show the disk angular momentum and the red lines show the planet spin axis. The disk evolution is similar to that found in hydrodynamical simulations \citep{MZA2020,Martin2021}, confirming that the one-dimensional wave-like linear model is at least a fair approximation to the full disk dynamics. The planet spin does not evolve in this case as there is no torque on the planet. This result for the disk evolution is independent of the disk mass. 

\section{Disk-planet evolution including planetary torque}

\begin{figure*} 
\center
\includegraphics[width=0.45\textwidth]{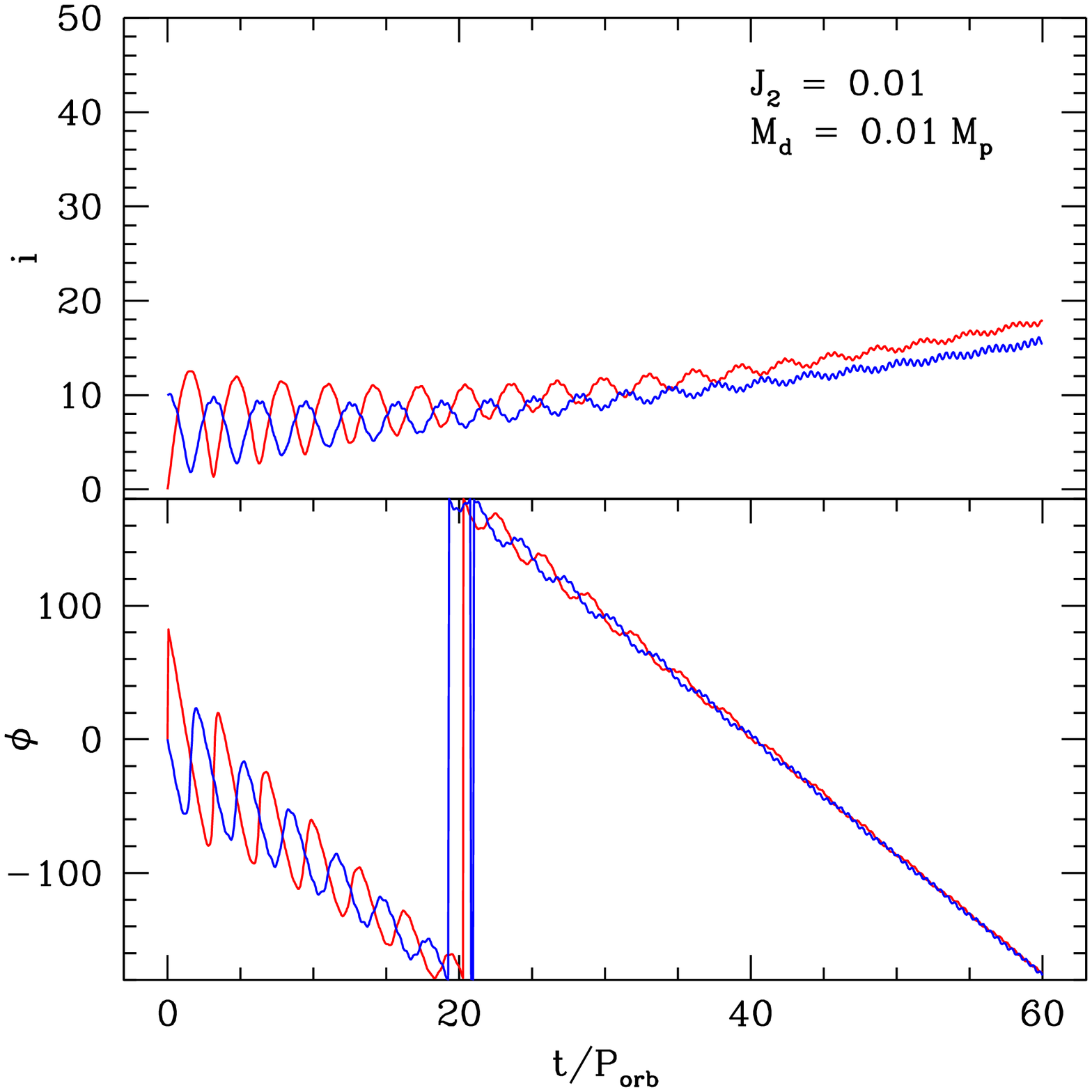}
\includegraphics[width=0.45\textwidth]{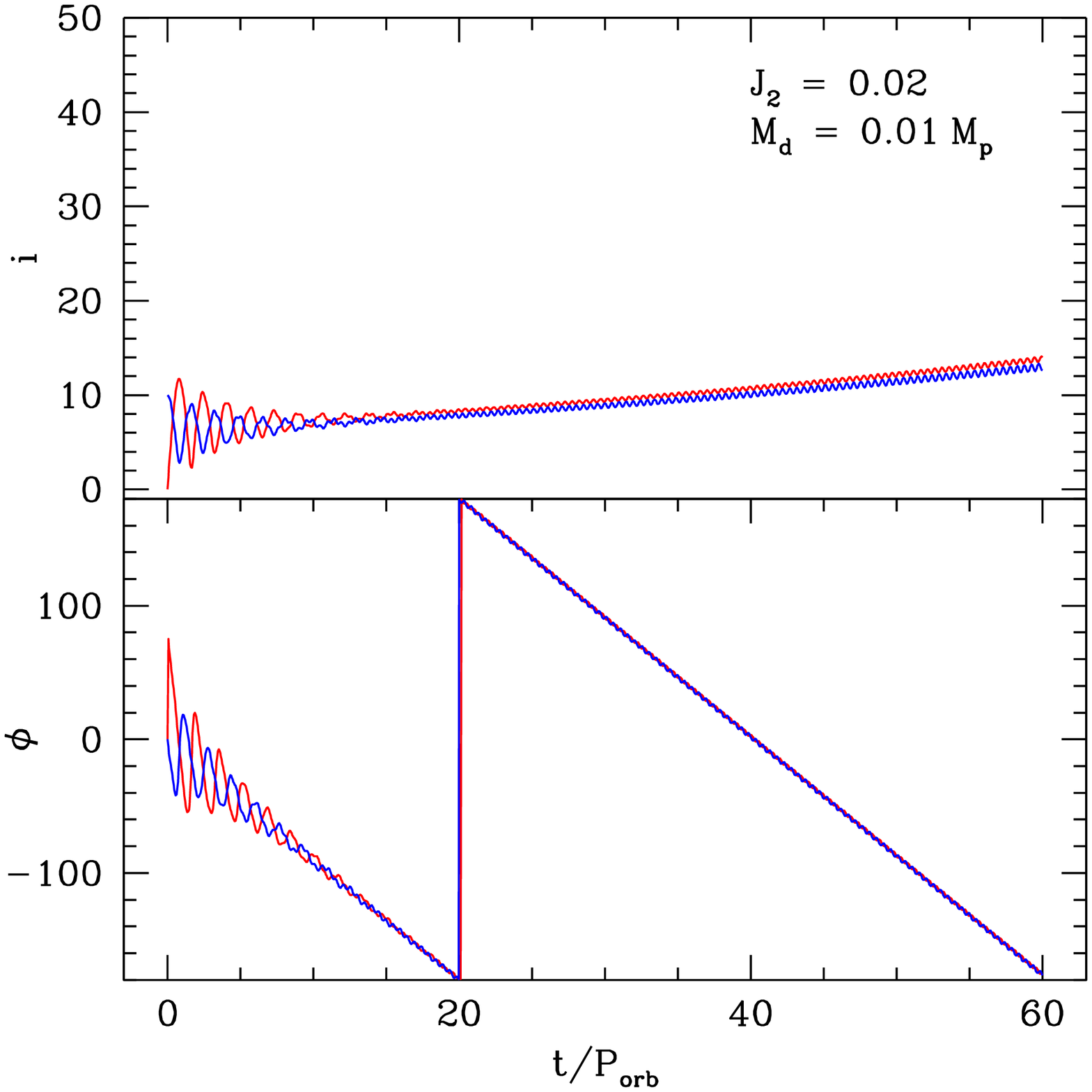}
\includegraphics[width=0.45\textwidth]{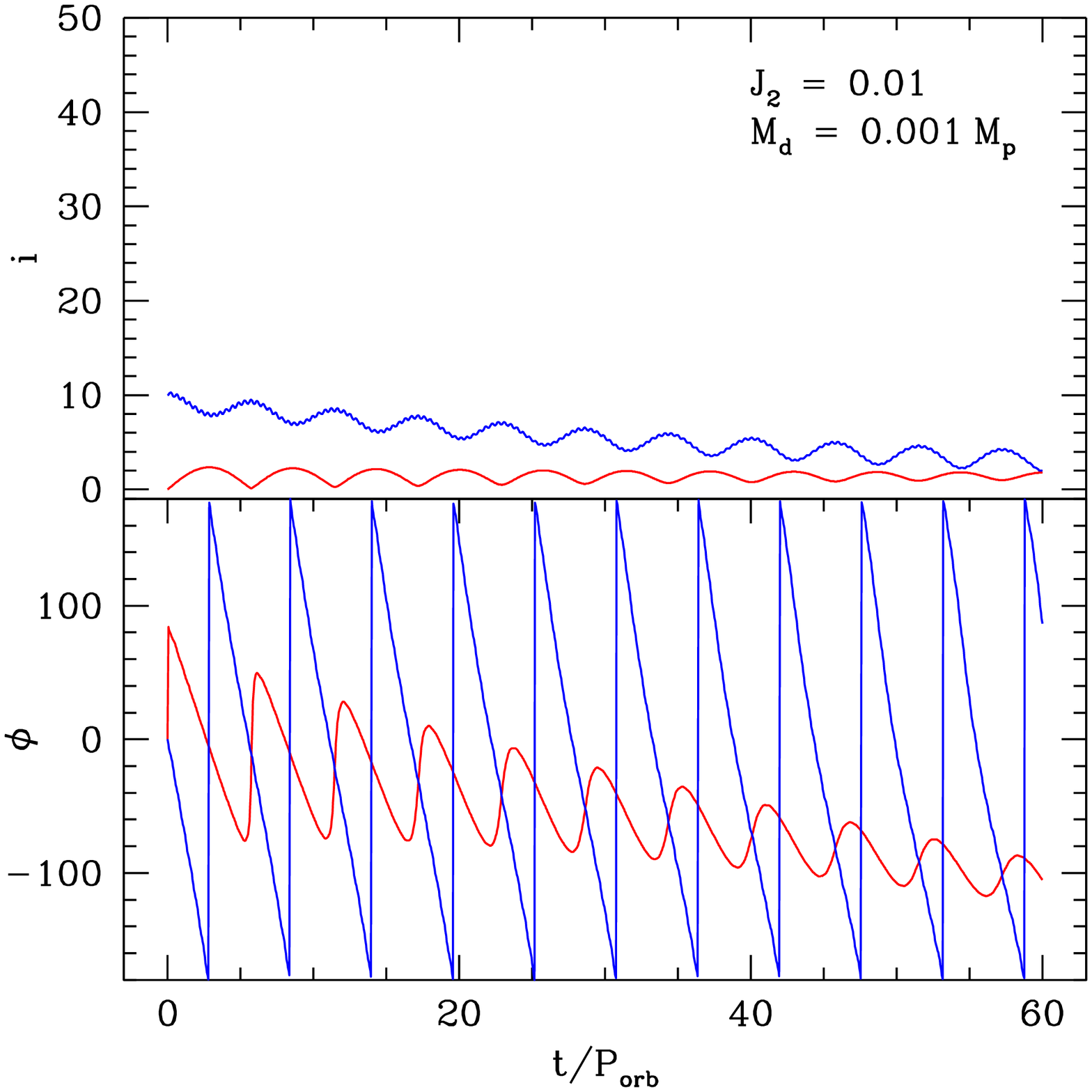}
\includegraphics[width=0.45\textwidth]{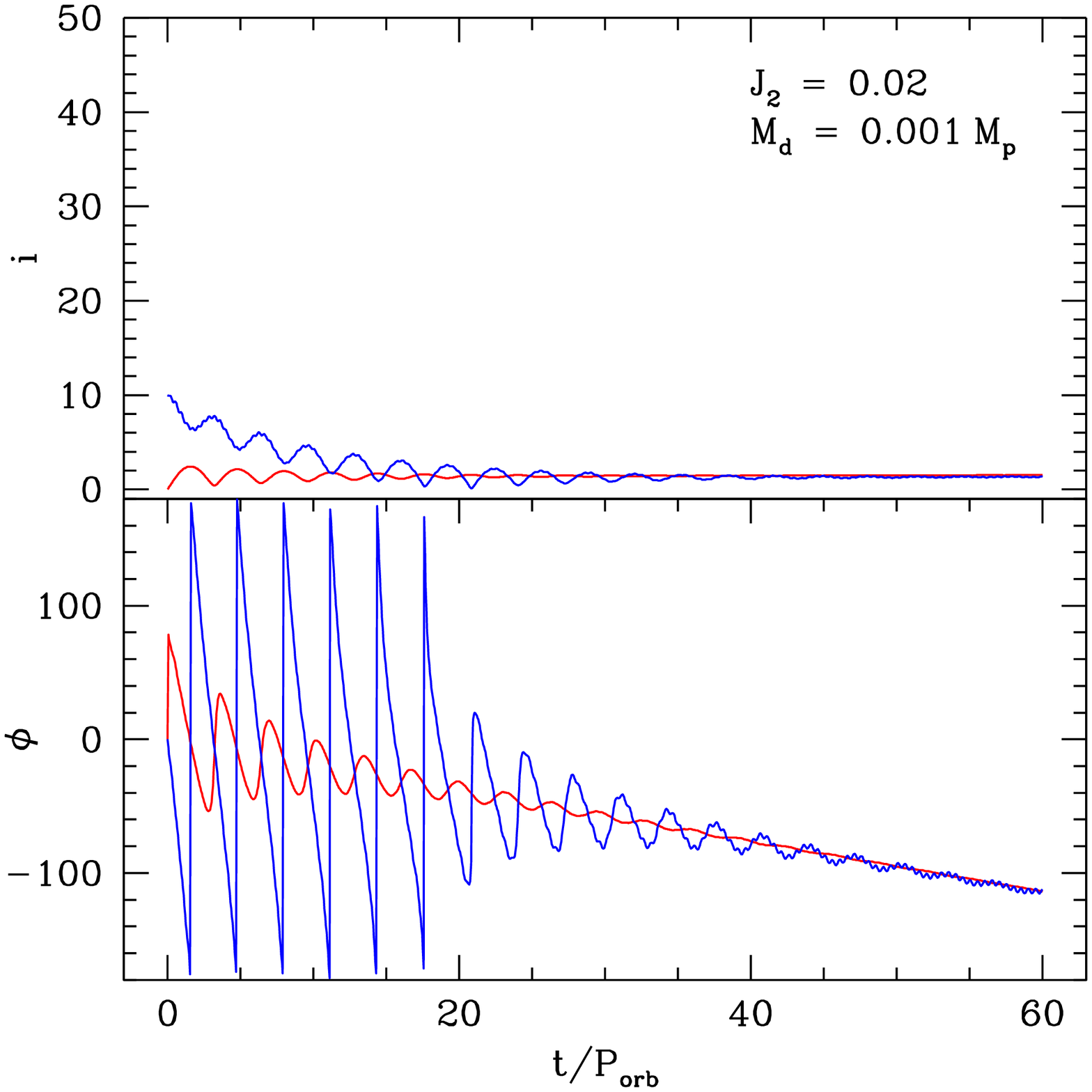}
\caption{Each panel is the same as Figure~\ref{j0} except $J_2=0.01$ (left panels), $J_2=0.02$ (right panels) and the disk mass is mass $M_{\rm d}=0.01\, M_{\rm p}$ (top panels) and $M_{\rm d}=0.001\, M_{\rm p}$ (bottom panels).   }
\label{figmain} 
\end{figure*}

We now consider two values for the quadrupole gravitational harmonic, $J_2=0.01$ and  $J_2=0.02$.  These $J_2$ values bracket those of Jupiter and Saturn, and are appropriate for planets whose rotation is rapid but well below the break-up value. The planet is assumed to rotate with period $P_{\rm rot}=10\,\rm hr$. Figure~\ref{figmain} shows the disk and planet evolution for two disk masses, $M_{\rm d}=0.01\,M_{\rm p}$ (upper panels), and $M_{\rm d}=0.001\,M_{\rm p}$ (lower panels), for both choices of $J_2$.  The planet and the disk precess around each other with damped inclination relative to each other. As these oscillations damp out, the stellar torque becomes the dominant effect. For the high mass disk, the planet and the disk both increase in inclination over time. However, for the low mass disks, the planet torque dominates the stellar torque and the disk and the planet do not increase significantly in inclination  over the timescales shown here. 
 There is a superposition of two modes: a damped mode in which the disk and the planet have oppositely directed tilts and a growing mode in which they have nearly equal tilts. In each case, the latter mode has a positive growth rate and eventually dominates but the growth rate is diminished by the coupling between the planet and the disk. Even in the low mass disc case, the planet obliquity increases in time over longer timescales than shown here.  

The timescale on which the inclination of the planet-disk system changes is longer than that of a disk without a planetary torque, but it is still relatively short, of the order of $10^2 \ P_{\rm orb}$. We do note that a high angular momentum disk is required to change the planet spin. Significant effects are therefore more likely to occur in planets that are farther from their host star since the circumplanetary disk is larger and therefore can have more angular momentum. 

\section{Disk-planet evolution with no stellar torque}

In order to make some analytic estimates we now  consider the timescale on which the planet spin axis changes as a result of the disk torque only in the absence of the star. We assume that the disk is flat but tilted so that  $\bm{l}_{\rm d}=\bm{l}_{\rm d}(t)$. The spin angular momentum of the planet is
\begin{equation}
J_{\rm planet}=2.5\times 10^{-7} 
 \left(\frac{M_{\rm p}}{10^{-3}\,\rm M_\odot}\right)
    \left(\frac{r_{\rm p}}{r_{\rm J}}\right)^{2}
  \left(\frac{P_{\rm rot}}{10\,\rm hr}\right)^{-1} 
\,\rm M_\odot au^2 yr^{-1},
\end{equation}
where $r_{\rm J}$ is the radius of Jupiter.
The angular momentum of the disk is
\begin{equation}
    J_{\rm disk}=\int_{r_{\rm in}}^{r_{\rm out}} 2 \pi r \Sigma (r^2 \Omega)\, dr
\end{equation}
and for typical parameters this is
\begin{align}
     J_{\rm disk}= ~~& 3.8\times 10^{-7}  
     \left(\frac{a_{\rm p}}{5.2\,\rm au}\right)^{1/2}
      \left(\frac{M_{\rm d}}{10^{-5}\,\rm M_\odot} \right) \cr & \times
      \left(\frac{M_{\rm p}}{10^{-3}\,\rm  M_\odot}\right)^{2/3}
        \left(\frac{M_{\rm s}}{\rm  M_\odot}\right)^{-1/6}
        \,\rm M_\odot au^2 yr^{-1}.
\end{align}
Note that this calculation depends on the surface density profile. We have taken a steep profile of $\Sigma \propto R^{-3/2}$. With $\Sigma \propto R^{-1/2}$, the disk angular momentum is larger by a factor of 1.5 for the same total disk mass.

Integrating equation~(1) over the disk we have 
\begin{equation} 
 J_{\rm disk}    \frac{\partial \bm{l}_{\rm d}}{\partial t}= T_{\rm total} \, \left(\bm{l}_{\rm d}\times \bm{l}_{\rm p}\right)
\label{disceq}
\end{equation}
where 
\begin{equation}
    T_{\rm total}=\int_{r_{\rm in}}^{r_{\rm out}} 2\pi r T\,dr.
\end{equation}
We can also write equation~(\ref{eqA}) as
\begin{equation} 
 J_{\rm planet}    \frac{\partial \bm{l}_{\rm p}}{\partial t}= -T_{\rm total} \, \left(\bm{l}_{\rm d}\times \bm{l}_{\rm p}\right).
\label{planeteq}
\end{equation}

\subsection{Planet-disk precession timescale}
With equations~(\ref{disceq}) and~(\ref{planeteq}) we see that the sum of the planet and the disc angular momentum is conserved. Thus, in the absence of the star, the planet and the disk both precess in a retrograde direction about the sum of the disk and planet angular momentum vector. The timescale for the precession is 
\begin{equation}
    t_{\rm prec}=2\pi f \frac{J_{\rm planet}}{T_{\rm total}} 
\end{equation}
where
\begin{equation}
    f=\frac{1}{\frac{J_{\rm planet}}{J_{\rm disk}}+1}. 
\end{equation}
The circumplanetary disk has a power law surface density profile $\Sigma \propto r^{-3/2}$ from $r_{\rm in}=2 r_{\rm p}$ up to $r_{\rm out}=0.4\,r_{\rm H}$. For typical parameters we find
\begin{align}
   \frac{ t_{\rm prec}}{P_{\rm orb}}\approx & \,\,3.2
   \left(\frac{r_{\rm p}}{r_{\rm J}}\right)^{5/2}
   \left( \frac{J_2}{0.01}\right)^{-1}
   \left(\frac{a_{\rm p}}{5.2\,\rm au}\right)^{-1}
   \left(\frac{M_{\rm d}}{10^{-5}\,\rm M_\odot} \right)^{-1}
   \\ \notag
   & \times 
   \left(\frac{M_{\rm p}}{0.001\,\rm M_\odot}\right)^{1/6}
    \left(\frac{M_{\rm s}}{\rm M_\odot}\right)^{1/3}
   \left(\frac{P_{\rm rot}}{10\,\rm hr}\right)^{-1} 
   \left(\frac{f}{0.6}\right) 
  .
\end{align}
This is in good agreement with the results shown in Figure~\ref{figmain}, since we find the precession timescales for the different models to be $3.2$ (model in the upper left panel), $1.6$ (upper right), $6.9$ (lower right) and $3.4\, P_{\rm orb}$ (lower right). We note that the factor $f$ depends upon the ratio of the disk to planet angular momenta.

\subsection{Planet-disk alignment timescale}
The alignment timescale between the disk and the planet spin is approximated by
\begin{equation}
    t_{\rm align}=\frac{(H/r)^2 \Omega_{\rm d}}{\alpha \omega_{\rm p}^2}
\end{equation}
\citep[e.g.][]{Bateetal2000,Lubow2018}, where $\Omega_{\rm d}=\Omega(r_{\rm out})$ and the precession rate is $\omega_{\rm p}=2\pi/t_{\rm prec}$. For typical parameters we find
\begin{align}
    \frac{t_{\rm align}}{P_{\rm orb}}= & ~~21.4 \,\ 
    \left(\frac{H/r}{0.14}\right)^2
    \left(\frac{\alpha}{0.01}\right)^{-1}
     \left(\frac{r_{\rm p}}{r_{\rm J}}\right)^{5}
   \left( \frac{J_2}{0.01}\right)^{-2}
   \left(\frac{a_{\rm p}}{5.2\,\rm au}\right)^{-2}
    \cr
   & \times 
   \left(\frac{M_{\rm d}}{10^{-5}\,M_\odot} \right)^{-2}
   \left(\frac{M_{\rm p}}{0.001\,\rm M_\odot}\right)^{1/3}
     \left(\frac{M_{\rm s}}{\rm M_\odot}\right)^{2/3}
   \left(\frac{P_{\rm rot}}{10\,\rm hr}\right)^{-2} \cr
   & \times \left(\frac{f}{0.6}\right)^2.
\end{align}
Since this is an exponential decay timescale, this is in rough agreement with the numerical simulations shown in Figure~\ref{figmain}. The alignment timescales are $21.4$ (upper left), $5.3$ (upper right), $100.6$ (lower left) and $25.2\,P_{\rm orb}$ (lower right). Note again that the factor $f$ depends upon the ratio of the disk to planet angular momenta.

\section{Discussion}
A circumplanetary disk that develops a tilt, due to the tidal tilt instability \citep{Lubow1992}, provides a substantial reservoir of misaligned angular momentum even when the disk mass is small relative to that of the planet. In this paper, we have shown that the disk misalignment can be communicated to the planet via the torques that result from planetary oblateness, leading to an efficient mechanism for generating obliquity. The mechanism will work (given our approximations) if, first, the circumplanetary disk properties (primarily the disk aspect ratio and size) allow for tilt growth. That tilt will change the planet's rotation state if, additionally, the disk mass is sufficiently large. 

Empirical constraints on circumplanetary disk masses are currently weak, while theoretical models span a broad range \citep{Lunine1982,Canup2002,Mosqueira2003,Lubow2013,batygin20}. Broadly speaking, static Minimum Mass Satellite Nebula models, those that include dead zones \citep{Gammie1996}, and those where the disk is a decretion rather than an accretion disk, lead to large circumplanetary disk masses that could exceed the threshold we have estimated for Jovian conditions. Simple viscous disk models, in which $\alpha \sim 10^{-2}$ and the planet is assembled in $\sim 10^6 \ {\rm yr}$, lead to sub-threshold masses. Since the size of circumplanetary disk is determined by the size of the Hill radius, which increases with distance from the star, all else being equal larger planetary semi-major axes lead to a greater likelihood of obliquity excitation. We note that the low accretion rate inferred from the H$\alpha$ luminosity of PDS~70b (Yifan Zhou, private communication), in the directly imaged PDS~70 system \citep{keppler18}, would imply a very low disk mass in the viscous scenario but could also be consistent with higher disk masses in the other models.

Our results are based upon describing the circumplanetary disk using one-dimensional equations, derived in the linear regime for wave-like warp propagation. The same tidal potential that is responsible for the tilt instability also drives spiral shocks in circumplanetary disks \citep{Zhu2016}, raising the question of whether a one-dimensional description is adequate. The tilt instability, however, is recovered in three-dimensional simulations using independent numerical methods (Smooth Particle Hydrodynamics and fixed-grid simulations), with a growth rate in approximate agreement with linear estimates \citep{MZA2020}. We therefore expect the current treatment to be adequate for moderate tilts, though simulations are indispensable for modelling very large misalignments. We have also ignored  the effects of accretion onto the circumplanetary disk \citep{Tanigawa12,Szulagyi2014,Schulik2020}. This is also an intrinsically three-dimensional effect, though we note that it could be approximately incorporated into our model using methods analogous to those used in dwarf nova disk models \citep{bath81}. 

Determining observationally whether the mechanism we have described here operates in real systems will clearly be difficult. As the Solar System example makes clear \citep{laskar93}, secular resonances in multiple planet systems can be efficient sources of late-time obliquity \citep[even Jupiter's low and presumably primordial obliquity is predicted to eventually increase;][]{saillenfest20}. Even when a system's current architecture is clearly non-resonant, excitation of obliquity could have occurred during a prior epoch of migration \citep{vokrouhlick15}. Young systems, observed during or shortly after the disk-embedded phase, provide the best opportunities. Our model suggests (and requires) that some  circumplanetary disks around giant planets \citep{Zhu2015b} ought to be found to be significantly tilted. In some cases the tilts may be large, approaching the critical misalignment for Kozai-Lidov oscillations \citep{Martin2021}. Planetary obliquities ought to be non-zero even for the youngest planets, and should be larger on average for planets at large orbital radii, which would have hosted more massive circumplanetary disks that are more efficient at communicating their misalignments to the planet. Smaller mass planets may also be favored, as their circumplanetary disks are expected to have a larger disc aspect ratio making them more unstable to tilting.

\section*{Acknowledgements} 
We thank Kaitlin Kratter and Zhaohuan Zhu for valuable discussions, and acknowledge support from NASA TCAN award 80NSSC19K0639. 
 
\bibliographystyle{mnras}

\end{document}